\documentclass{article}

\usepackage{PRIMEarxiv}

\usepackage[utf8]{inputenc} 
\usepackage[T1]{fontenc}    
\usepackage{hyperref}       
\usepackage{url}            
\usepackage{booktabs}       
\usepackage{amsfonts}       
\usepackage{amsmath,bm}
\usepackage{nicefrac}       
\usepackage{microtype}      
\usepackage{lipsum}
\usepackage{fancyhdr}       
\usepackage{graphicx}       
\usepackage{rotating,tabularx}
\usepackage{multirow}

\newcommand{\expec}{\operatorname{E}}
\newcommand{\prob}{\operatorname{P}}

\newcommand\numberthis{\addtocounter{equation}{1}\tag{\theequation}}

\graphicspath{{media/}}     

\title{Individual claims reserving using activation patterns
\thanks{\textit{\underline{Citation}}: 
\textbf{Michaelides, M., Pigeon, M. \& Cossette, H., Individual Claims Reserving using Activation Patterns \textit{European Actuarial Journal}. (2023). https://doi.org/10.1007/s13385-023-00355-3} 
}}

\author{
  Marie Michaelides\\
  Chaire Co-operators en analyse des risques actuariels \\
  Département de Mathématiques \\
  Université du Québec à Montréal \\
  Montréal\\
  \texttt{michaelides.marie@uqam.ca} \\
  \And
  Mathieu Pigeon \\
  Chaire Co-operators en analyse des risques actuariels \\
  Département de Mathématiques \\
  Université du Québec à Montréal \\
  Montréal\\
   \And
  Hélène Cossette \\
  École d'actuariat \\
  Université Laval \\
  Québec\\
}

\begin{document}
\maketitle

\begin{abstract}
A claim often impacts not one but multiple insurance coverages provided in the contract. To account for this multivariate feature, we propose a new individual claim reserving model built around the activation of the different coverages to predict the reserve amounts. Using the framework of multinomial logistic regression, we model the activation of the different insurance coverages for each claim and their development in the following years, i.e., the activation of other coverages in the later years and all the possible payments that might result from them. As such, the model allows us to complete the individual development of the open claims in the portfolio. Using a recent automobile dataset from a major Canadian insurance company, we demonstrate that this approach generates accurate predictions of the total reserves and the reserves per insurance coverage. This analysis allows the insurer to get new insights into the dynamics of his claims reserves.
\end{abstract}

\keywords{Granular reserving \and Dependence modelling \and Insurance coverages \and Multinomial}

\section{Introduction}\label{sec1}

Claims reserving is known to be one of the most crucial tasks performed by actuaries in insurance companies all over the world. Insurers must accurately predict future liabilities related to open claims. It allows them to answer the reporting standards they are subject to and to preserve sufficient capital with which they can set aside adequate reserves to fulfil their obligations to the policyholders and avoid financial ruin. 

In actuarial practice, claims data is commonly aggregated on an occurrence year and development year basis in run-off triangles. These then help actuaries to evaluate the reserves for the portfolio as a whole. One ubiquitous method that uses such triangles is the Chain Ladder model introduced by Mack \cite{mack1993} and further discussed in \cite{mack1994}, \cite{mack1999} or \cite{mack2000}. 

Over the years, several authors have challenged the robustness of this model. In particular, the recent increase in the quantity and availability of data has contributed to questioning the use of such aggregate methods. The early work of Buhlmann et al. \cite{buhlmann1980}, Hachemeister \cite{hachemeister1980}, and Norberg \cite{norberg1986} attempted to benefit from these larger quantities of data. However, a few more years were required to obtain the necessary computing resources to move from the classical run-off triangles to the so-called micro-level claims reserving models. Antonio and Plat \cite{antonio2014} were the first to truly incorporate information related to the policyholder or even the claim itself into their model by building on the work mentioned above as well as on prior work performed by Norberg \cite{norberg1993}, \cite{norberg1999} and Haastrup and Arjas \cite{haastrup1996}. They demonstrate that using the detailed information available to the insurer at the claim level allows them to obtain more accurate predictions for the reserves. Following their lead, Pigeon et al. \cite{pigeon2013} also propose a fully parametric discrete-time model for the payments and then extend their work in \cite{pigeon2014} to include incurred losses as well. Many authors have since then contributed to this area of research which is still very active today, as can be seen, for example, from the recent contribution of \cite{crevecoeur2022} who focus on the treatment of RBNS claims.

Other authors have also opened the way to non-parametric approaches to claims reserving. Wüthrich \cite{wuthrich2018} was the first to introduce the use of Breiman et al. \cite{breiman1984}'s Classification and Regression Tree (CART) algorithm in a micro-level reserving model. Building on this work, Lopez et al. (\cite{lopez2019}, \cite{lopez2016}) apply the CART algorithm with censored data using, respectively, survival analysis and copulas to account for the possible dependence between the development time of the claim and its ultimate amount. More recently, \cite{delong2020} and \cite{delong2020b} propose to use neural networks to model the joint development of individual payments and claims incurred, thus benefiting from individual claims data in a collective reserving framework.

In addition to the constant increase in the quantity of data, the growing diversification of the products offered by insurers contributes to further complexifying the work of actuaries. Evolving in a very competitive world and taking advantage of the rise of new technologies, insurers must constantly remain aware of changes and evolutions in their clients' needs. To answer them, they often diversify their offer, multiplying coverages provided within a policy. Actuaries must refine their models to keep up with this diversity in their portfolios. To do so, they must consider the correlation between these different risks. Although some authors have already contributed to the claims reserving literature with models that include diverse forms of dependence, the specific inter-coverage dependence that we highlight in this paper has yet to be modelled in such a context, to the best of our knowledge. Among others, Zhou and Zhao \cite{zhou2010} and Lopez \cite{lopez2019} used copulas to model the dependence between the event times and delay in the development of a claim or the development time and the final amount of the claim. Pe$\Tilde{\text{s}}$ta and Okhrin \cite{pesta2014} use time series and copulas to consider the dependence between payment amounts made at different stages in the development of a claim. 

In the pricing literature, we find examples of joint modelling for correlated risks within a portfolio. In particular, Frees, Valdez \cite{frees2008}, and Shi \cite{frees2009} used copulas to model the dependence between different claim types in automobile insurance. They begin by identifying the coverage(s) impacted by a claim, then those for which a payment is made, before predicting the associated severity. Also, with automobile insurance data, Shi et al. \cite{shi2016} account for the cross-sectional and temporal dependence among multilevel claims with a copula regression for multivariate longitudinal claims. Closer to what we intend to do in this paper, Shi and Kun \cite{shi2021} use a multinomial logistic model for the occurrence of correlated risks. Yang and Shi \cite{yang2019} apply a methodology similar to that of Shi et al. \cite{shi2016} and model multilevel property insurance claims with a Gaussian copula while considering zero inflation. Frees et al. \cite{frees2010} model the frequency and severity of multilevel property claims with, respectively, logistic regressions and Gaussian copulas. Finally, Frees et al. \cite{frees2013} employ a similar approach for healthcare insurance multilevel claims and use multivariate binary regressions to model the different payment types, then Gaussian copulas again for the severities. More recently, Côté et al. \cite{cote2022} extend the work of Frees and Valdez \cite{frees2008} and Frees, Valdez and Shi \cite{frees2009} by introducing a Bayesian model for multivariate and multilevel claim amounts, therefore facilitating the treatment of open claims which are of crucial importance in insurers' datasets. 

This paper analyses a recent automobile dataset from a major Canadian insurer that provides multiple insurance coverages for each policy. Our goal is two-fold: first, we want to account for the multilevel feature of insurance claims in a reserving framework and model the dependence between the different coverages included within a policy of the portfolio at hand. Second, we seek to separately model the main events that take place during the lifetime of a claim, namely the activation of a coverage and the time of payment which we consider as two separate events. As such, we contribute to the efforts to predict the development of the claims already illustrated in \cite{crevecoeur2023}. We begin in Section \ref{sec:data} by introducing the dataset. Section \ref{sec:Model} then presents the statistical model for reported but not settled (RBNS) claims based on activation patterns in which dependence between insurance coverages is captured via a multinomial logistic regression. In Section \ref{sec:application}, we present the model's results applied to the Canadian dataset and compare them to the results of two additional models. We conclude our analysis in Section \ref{sec:conclusion}.

\section{Data}
\label{sec:data}
This section presents an overview and some exploratory analysis of our dataset.

Our data contain 656,153 automobile insurance claims which have occurred between the $1^{\text{st}}$ of January 2015 and the $30^{\text{th}}$ of June 2021 in Canada or the United States. Each of these can impact one of four different insurance coverages provided for each policy by the insurer. In addition, we have information related either to the insured, the car driven, or the claim itself.

\begin{figure}[h!]
\centering
\includegraphics[width=1\textwidth]{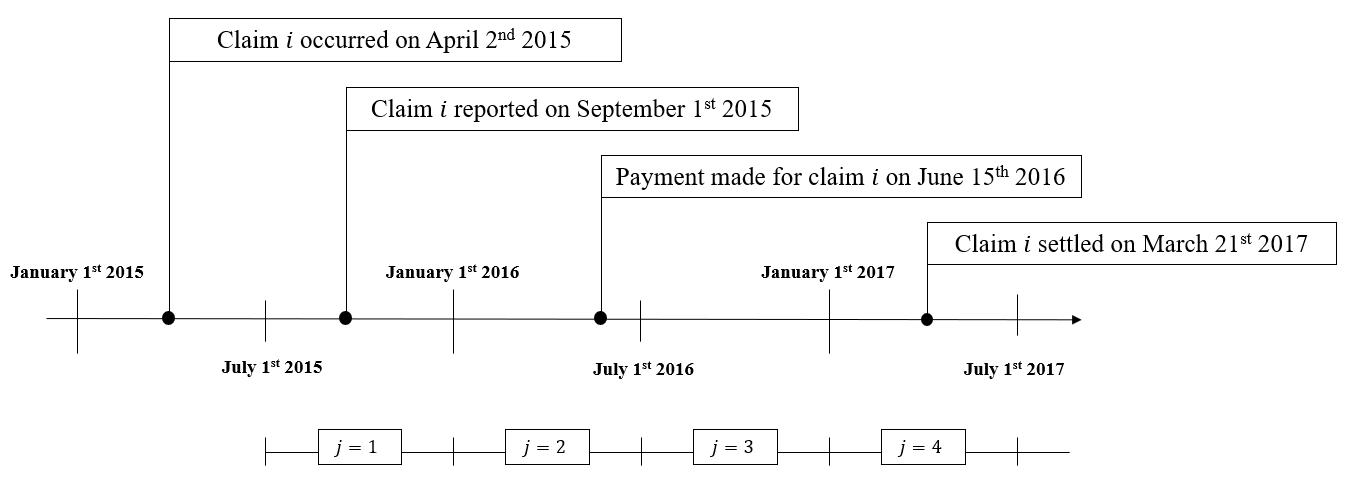}
\caption{Data grouping in periods of six months. In this illustration, claim $i$ is reported on September $1^{\text{st}}$ 2015 after a delay of one period since its occurrence. It is then in its first ($j=1$) development period. Payment is made for that claim in development period $j=2$. The claim settles on March $21^{\text{st}}$ 2017, i.e., in its $4^{\text{th}}$ development period.}
\label{fig:Fig1}
\end{figure}

Note that, as is typically done in claims reserving, we work in a discrete-time framework and group the data in distinct development periods. Due to the limited number of calendar years available in the data and the short-tailed nature of the portfolio, we choose to split each calendar year into two periods of six months and use these as a basis for development periods. We illustrate this in Figure~\ref{fig:Fig1}.

Using half-years as time steps will allow us, in the rest of this paper, to gain a more refined understanding of the development of claims throughout their lifetime. However, because academics and practitioners typically use one-year rather than six-month time steps in claims reserving, we will ensure that the model we develop in this paper for this specific portfolio remains general enough to be easily applied to other time units.

We first explore the four insurance coverages the insurance company offers for each policy in force before taking a closer look at the risk factors we will include in the analysis in Section~\ref{sec:Model}.

\subsection{Insurance coverages}
Each of the 656,153 claims in our dataset impacts one or up to four coverages the insurer offers to the policyholders: the Accident Benefits coverage (loss of revenue, funeral expenses for the insured), the Bodily Injury coverage (loss revenue or medical expenses for a third party), the Vehicle Damage coverage (damages incurred to the insured's or a third party's vehicle) and the Loss of Use coverage (temporary replacement vehicle and alternative means of transportation).

To better understand the dynamics of our portfolio, Table~\ref{tab:covweights} shows the importance of each coverage in terms of the proportion of claims, total cost, and total reserve, taking January $1^{\text{st}}$ 2019 as the valuation date. Unsurprisingly, the Vehicle Damage coverage is the most frequently observed, with $96.39\%$ of the 656,153 claims activating it, while only $5.70\%$ of the claims trigger the Bodily Injury coverage.  

The proportions are slightly different when we look at the repartition of the total portfolio cost over the four coverages. In particular, Loss of Use is the coverage that weights the least cost-wise, representing less than $4\%$ of the total cost of the portfolio. Despite being activated by less than $6\%$ of all claims, the Bodily Injury coverage still represents around $13\%$ of the total portfolio cost. However, the most striking difference arises from the proportion of the total reserve attributed to each coverage. We calculate the percentages in the last column of Table \ref{tab:covweights} with the $1^{\text{st}}$ of January 2019 chosen as valuation date. Even though only $15\%$ of all claims activate the Accident Benefits and/or Bodily Injury coverages, they amount to $85\%$ of the total reserve. In comparison, the Loss of Use coverage activated by more than half the claims represents less than $1\%$ of the total reserve.
\begin{table}[h]
\begin{center}
\caption{Weight of each coverage in the portfolio. The percentages with respect to the total reserve are calculated by taking the $1^{\text{st}}$ of January 2019 as valuation date}
\label{tab:covweights}
\begin{tabular}{@{}lrrr@{}}
\toprule
  Coverage & \multicolumn{1}{c}{$\%$ of claims} & \multicolumn{1}{c}{$\%$ of total cost} & \multicolumn{1}{c}{$\%$ of total reserve}\\  
\midrule
Accident Benefits  & 9.42 & 12.82 & 30\\
 Bodily Injury  & 5.70 & 13.13 & 55 \\
 Vehicle Damage  & 96.39 & 70.44 & 14\\
 Loss of Use  & 51.89 & 3.61 & $<1$\\
 \bottomrule
\end{tabular}
\end{center}
\end{table}

Table \ref{tab:infocov} shows descriptive statistics related to the payments made for each coverage. Bodily Injury claims are those with the largest average payments, while the Loss of Use coverage presents the lowest one, as was already hinted at in Table \ref{tab:covweights}. The Accident Benefits, Bodily Injury, and Vehicle Damage coverages present large values for the payments in the higher quantiles, indicating that their corresponding distributions are probably heavy-tailed. 
\begin{table}[h]
\begin{center}
\caption{Descriptive statistics for the severity of payments of the four insurance coverages}
\label{tab:infocov}
\begin{tabular}{@{}l r r r r r r r@{}}
\toprule
\multirow{2}{*}{Coverage}& \multicolumn{1}{c}{\multirow{2}{*}{Mean}} & \multicolumn{1}{c}{\multirow{2}{*}{Std. dev.}} & \multicolumn{4}{c}{Quantiles} & \multicolumn{1}{c}{\multirow{2}{*}{Max.}} \\\cline{4-7}
&  &  & \multicolumn{1}{c}{0.5} & \multicolumn{1}{c}{0.75} & \multicolumn{1}{c}{0.95} & \multicolumn{1}{c}{0.99} & \\
\midrule
Accident Benefits & 12,386 & 53,561 & 3,215 & 6,909 & 47,757 & 127,896 & 2,435,334 \\
Bodily Injury & 23,271 & 76,027 & 4,000 & 15,150 & 98,449 & 322,612 & 2,039,570\\
Vehicle Damage & 5,040 & 8,121 & 2,605 & 5,830 &  17,984 & 40,611 & 149,399 \\
Loss of Use & 545 & 620 & 419 & 714 & 1,000 &  2,336 & 52,777 \\
\bottomrule
\end{tabular}
\end{center}
\end{table}

\paragraph{Activation and payment delays}
We define each insurance coverage's activation and payment delays. The activation delay corresponds to the delay (in period units) between the reporting date of the claim and the date the insurer first realises that this claim triggers the coverage. The payment delay is the time (in period units) between the activation date of a coverage and the date the insurer makes the first payment towards the claim related to that specific coverage. Note that we assume here that the activation of a coverage does not necessarily imply a payment. As means of illustration, consider again the example provided in Figure \ref{fig:Fig1} and suppose that claim $i$ only activates the Accident Benefits coverage directly upon reporting. Since the activation of the coverage takes place in the same development period as the reporting date, the activation delay for that coverage is equal to 0. The first payment is then made in development period $j=2$, indicating a payment delay equal to 1 for the Accident Benefits coverage triggered by claim $i$.

Table \ref{tab:actidelays} displays the average activation delay per coverage. The Accident Benefits, Vehicle Damage, and Loss of Use coverages are typically activated in the same period as the reporting date for the majority of claims (corresponding to the \textit{No delay} column). However, for the Bodily Injury coverage, we observe that around $15\%$ of the claims activate it with a delay of at least one development period after the reporting date.  

Appendix \ref{ap:observed_acti} further illustrates the dynamics between the different coverages by presenting the activation patterns observed in the dataset for the first development period of the claims. 
\begin{table}[h]
\begin{center}
\caption{Percentage of claims with different activation delays for the four coverages}
\label{tab:actidelays}
\begin{tabular}{@{}l r r r r r@{}}
\toprule
& \multicolumn{5}{c}{Activation delays} \\
\midrule
  Coverage & \multicolumn{1}{c}{No delay} & \multicolumn{1}{c}{1 period} & \multicolumn{1}{c}{2 periods } & \multicolumn{1}{c}{3 periods } & \multicolumn{1}{c}{$\geq 4$ periods } \\ 
 \midrule
 Accident Benefits  & 93.84 & 5.73 & 0.29 & 0.08 & 0.06\\
 Bodily Injury  & 85.86 & 9.86 & 1.46 & 1.13 & 1.69\\
 Vehicle Damage  & 94.14 & 5.65 & 0.13 & 0.04 & 0.04 \\
 Loss of Use  & 92.10 & 7.73 & 0.13 & 0.03 & 0.01\\
 \bottomrule
\end{tabular}
\end{center}
\end{table}

\begin{figure}[h!]
\centering
\includegraphics[width=1\textwidth]{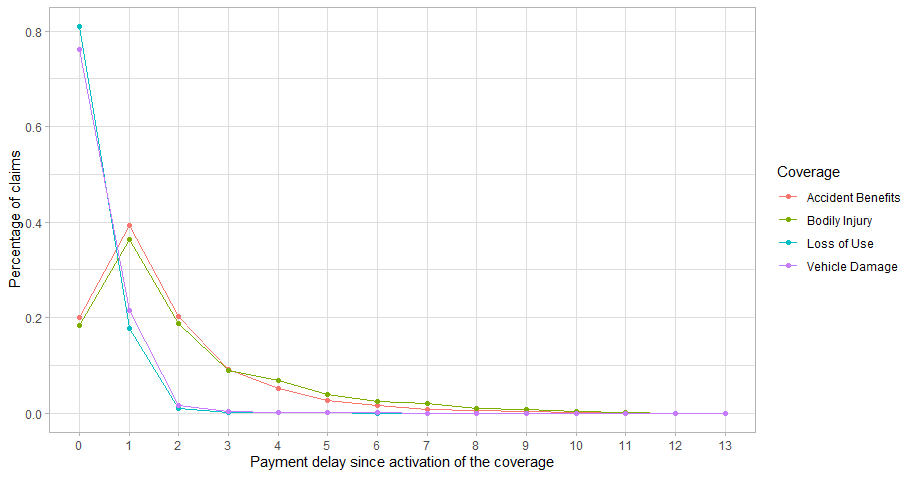}
\caption{Payment delays per coverage. A delay of 0 period means that the payment occurs in the same period the coverage becomes active}
\label{fig:Fig2}
\end{figure}

Once a claim activates a coverage, we are interested in knowing when the first payment occurs. The payment delay, illustrated for our data in Figure~\ref{fig:Fig2}, refers to the time (in six-month units) that elapsed between the moment a claim activates a coverage and the date the insurer records a first payment for that same coverage. We see from Figure~\ref{fig:Fig2} that the Accident Benefits and Bodily Injury coverages, on the one hand, and the Vehicle Damage and Loss of Use coverages, on the other hand, display very similar shapes. The Loss of Use coverage is the one for which payments occur the fastest after activation, with over $80\%$ of the claims related to that coverage receiving a payment in the same development period as the activation period. Almost all Loss of Use claims and Vehicle Damage claims will have received their first payment by the end of the second development period after activation of that coverage. Conversely, the Bodily Injury coverage is the one for which time to the first payment is the longest. Only around $20\%$ of the claims that activate that coverage receive a first payment in the same development period as the activation of the coverage. Several Bodily Injury claims will receive their first payment only after the $8^{\text{th}}$ development period. The situation for the Accident Benefits coverage is very similar.

\subsection{Risk factors}
\label{sec:riskfactors}
Table \ref{tab:riskfactors} presents the risk factors that we will use to build the reserving model. We provide further insights on these covariates in Appendix \ref{ap:riskfactors}. Note that the variables \texttt{GENDER} and \texttt{YOB} contained a significant proportion of missing values. We used the \texttt{R} package \texttt{mice} to fill them in. This package uses Fully Conditional Specification (FCS) where binary data (\texttt{GENDER}) and unordered categorical data (\texttt{YOB}) are imputed using, respectively, logistic and polytomous logistic, regressions (see \cite{mice2011} for more details). Note that this data-filling procedure did not cause any significant changes in the results we will discuss later. 
\begin{table}[h!]
\begin{center}
\caption{Description of risk factors}\label{tab:riskfactors}
\begin{tabular}{@{}p{3cm}p{13cm}@{}}
\toprule
 \multicolumn{2}{l}{Risk factors}\\   
\midrule
 GENDER  & Gender of the insured. \\
 YOB  & Year of birth: decade during which the insured was born. \\
 VU & Use of the vehicle made by the insured. \\
 AM & Annual distance driven (in km). \\
 PROV  & Place of occurrence of the claim: one of the Canadian provinces or the USA. \\
 FR  & Fault rating: evaluation of the insured's level of responsibility in the accident. \\
 REP\_DELAY & Reporting delay (in 6-months periods) between the occurrence and reporting dates of the claim. \\
 ACT\_DELAY & Activation delay per coverage (in 6-month periods) between the reporting date of a claim and the first date at which the insurer realises the claim triggers the corresponding coverage.  \\
 PAY\_DELAY & Delay (in 6-month periods) between the activation date of a coverage by a claim and the date at which a first payment is made towards that claim relating to that specific coverage.\\
\bottomrule
\end{tabular}
\end{center}
\end{table}

\section{An activation pattern model for claims reserving}
\label{sec:Model}
This section introduces the model we built to predict the granular reserves for the portfolio of claims described in Section~\ref{sec:data} considering the dependence between the insurance coverages. Note that even though we tailor the model to the specific needs of our dataset, our goal is to provide the most general description of the model so that an interested party may easily apply it to other studies.

We first specify the notation and the model components before presenting the model more formally for different development periods and proposing a simulation routine. 

\begin{figure}[h!]
\centering
\includegraphics[width=0.9\textwidth]{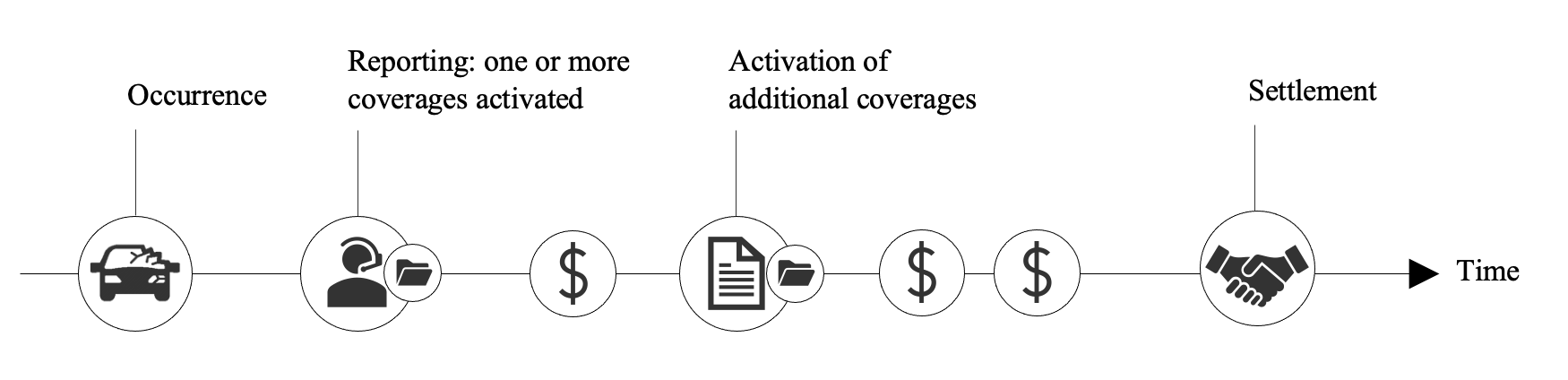}
\caption{Development process of a claim}
\label{fig:Fig3}
\end{figure}

Figure \ref{fig:Fig3} shows the typical development process for a single claim. When a claim occurs, the policyholder reports it to the insurance company. As we explained in Section~\ref{sec:data}, this can happen with or without delay. Once reported, the insurer records the claim and opens a new file in his claims management system. The reporting of the claim activates at least one of the coverages provided with the policy. The insurer can then start making payments to the policyholder. During its development, new information related to the claim can be brought to the insurer, which can result in the activation of one or more additional coverages. Then the insurance company will continue to make payments until the settlement of the claim. 

 We seek to model how a claim can activate multiple coverages upon reporting or later and the underlying dependence between them.  

\subsection{Notation}
\label{sec:notation}
Let $c$ with $c=1, \ldots, C$ denote a coverage provided within the policy by our insurer. Each policy in force can incur a claim that will impact one or more of the $C$ insurance coverages that the policyholder benefits from. For each claim $i$, with $i=1, \ldots, n$ and development period $j$ with $j=1, \ldots, J$, we define the following components: 
\begin{itemize}
\item  $\bm{A}_{i,j}$ is a $1 \times C$ random vector whose entries $A_{i,j,c}$ each take the value $0$ or $1$, i.e.,  $\bm{A}_{i,j} \in \{0,1\}^C$. We call  $\bm{A}_{i,j}$ the activation pattern vector that indicates which of the $C$ coverages claim $i$ activates in its $j^{th}$ development period. Given $C$ insurance coverages and assuming that at least one must be activated when claim $i$ is reported, there are $V=2^{C}-1$ different activation pattern vectors in development period $j$. We denote $\mathcal{V}$ the set of these $V$ patterns. Note that at this stage, we do not assume any constraints on the evolution of $\bm{A}_{i,j}$ over time in order to keep the definition as general as possible. Further details will however be added for the sake of our application to the dataset at hand in Section \ref{sec:j_plus_k}.\\  
\item $(P_{i,j,c} \vert A_{i,j,c}=1) \in\{0,1\} $ is a random variable that takes the value $1$ if, for the corresponding coverage $c$, the insurer makes at least one payment for claim $i$ in development period $j$, given that this coverage is active, or the value 0 otherwise. It is undefined if $A_{i,j,c}=0$. We call $P_{i,j,c}$ the payment pattern that depends on the random vector $\bm{A}_{i,j}$. \\
\item $(Y_{i,j,c}\vert P_{i,j,c}=1 ) \in \mathbb{R}^+$ is a positive continuous random variable that represents the severity associated with coverage $c$ when payments have been made for it for claim $i$ in development period $j$. $Y_{i,j,c}$ is thus not defined if the corresponding $P_{i,j,c}$ is equal to 0, i.e., if there was no payment for coverage $c$ and claim $i$ in its $j^{th}$ development period. For claims with longer development delays, we also define the random variable $\Tilde{Y}_{i,j,c} \in \mathbb{R}^+$ that represents the total remaining severity of claim $i$ for coverage $c$, from development period $j$ until settlement of the claim. We will discuss these longer claims in Section \ref{sec:j_star_plus_k}.
\item $\bm{x}_{i}'$ is a $1\times m$ vector of covariates for claim $i$ that contains static risk factors. 
\end{itemize}

\begin{figure}[h!]
\centering
\includegraphics[width=1\textwidth]{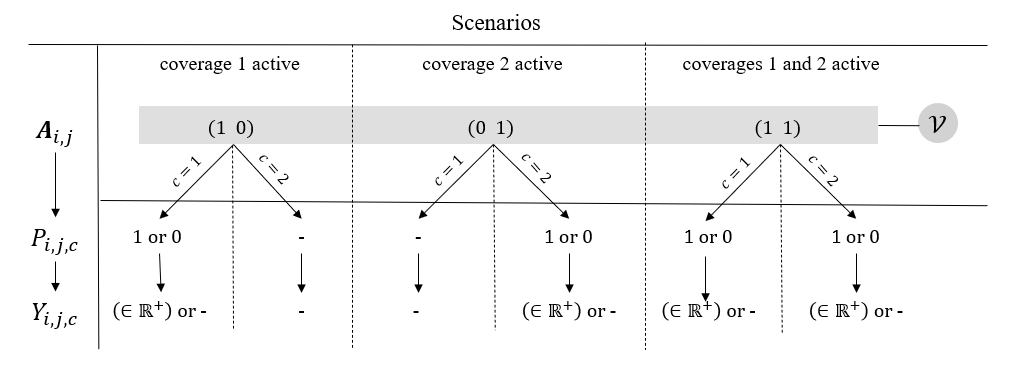}
\caption{Illustration of the possible scenarios with $C=2$ insurance coverages. The symbol ``-'' stands for the cases where the patterns are not defined.}
\label{fig:Fig4}
\end{figure}

\noindent
\small \textit{Example 1. To illustrate the notation, we consider a simple example with $C=2$ different insurance coverages. The set of possible activation patterns is $\mathcal{V} = \{(1\hspace{0.1cm} 0) \hspace{0.2cm} (0\hspace{0.1cm} 1) \hspace{0.2cm} (1 \hspace{0.1cm}1) \}$, corresponding respectively to the scenarios where a claim~$i$ activates either only the first coverage, only the second coverage or both of them in the same development period $j$. For each of these three scenarios, various realisations of $(P_{i,j,c} \vert  A_{i,j,c}=1)$ and $(Y_{i,j,c} \vert P_{i,j,c}=1)$ are possible, as illustrated in Figure \ref{fig:Fig4}.}
\normalsize

\subsection{Statistical model}
\label{sec:model}
Using the notation introduced in Section \ref{sec:notation}, we focus now on presenting our model for various development periods. We begin in the first development period before moving on to the following ones up to period $j^*$. We then present what happens after period $j^*$ for claims with longer settlement delays. In our notation, the insurer chooses $j^*$ as the development period from which the activation pattern for a claim stabilises, i.e., the last development period in which the claim might still activate additional coverages. This \textit{stabilisation point} in the activation of coverages may arise from, among others, the nature of the claims (automobile insurance claims typically present very short settlement delays) or even the legislative system in which the insurer evolves (some types of claims may not be pursued if a certain period has passed since occurrence). We finally describe the severity models used for each insurance coverage.
 
Knowing the number of claims reported in any given period, our model allows us to determine which coverages these claims activate and to predict their future development until development period $j^*$. Thus, we could apply the model to reported but not paid (RBNP) claims and reported but not settled (RBNS) claims. The model does not provide a complete framework for predicting incurred but not reported (IBNR) claims. It lacks a process predicting the number and characteristics of the IBNR claims. However, such processes have been widely discussed in the literature (see, for example, Antonio and Plat (2014) \cite{antonio2014} where the authors use a Position Dependent Marked Poisson Process) and can easily be used as a preliminary step to the model presented here. Once the insurer knows the number of IBNR claims, he may use the model to estimate their development from the first period onward.

\subsubsection{Development period $j=1$}
\label{sec:j}
For claim $i$ in development period $j=1$ and given $C$ insurance coverages, there exists $V_0=2^{C}-1$ possible realisations of the random activation pattern vector $\bm{A}_{i,1}$, all included within the set $\mathcal{V}_0$. This is the initial set of all possible activation patterns once a claim has entered the reporting system of the insurer. We write these realisations $\bm{v}$ and use a multinomial logit regression to model the probability that $\bm{A}_{i,1}$ takes one of them:
\begin{align}
    \prob[\bm{A}_{i,1} = \bm{v}] =
    \frac{\exp{(\bm{x}_{i}'\bm{\beta}_{1,v})}}{\sum_{\nu \in \mathcal{V_{0}}} \exp{(\bm{x}_{i}' \bm{\beta}_{1,v}})},
    \label{eq:logistic}
\end{align}
where $\bm{x}_{i}'$ is the vector of covariates introduced in Section \ref{sec:notation}. $\bm{\beta}_{1,v}$ is the $1\times m$ vector of parameters that can vary with the different coverages and thus depends on the $v^{th}$ pattern. 

Knowing which coverages claim $i$ activates in the first development period thanks to $\bm{A}_{i,1}$, the insurer can move on to the next step of the claim development process depicted in Figure \ref{fig:Fig1} and determine which of the active coverages will incur a payment within the period. For each insurance coverage $c$, we assume that
\begin{align}
   (P_{i,1,c} \vert A_{i,1,c}=1)  \sim \text{Bernoulli}\big(\pi_{1,c}(\bm{x}_{i}, \bm{\gamma}_{1,c})\big).
\end{align}
For development period $j=1$, the probability $\pi_{1,c}(\bm{x}_{i}, \bm{\gamma}_{1,c})$ is given by
\begin{align*}
    \pi_{1,c}(\bm{x}_{i}, \bm{\gamma}_{1,c}) = \frac{\exp{(\bm{x}_{i}'\bm{\gamma}_{1,c})}}{1+ \exp{(\bm{x}_{i}' \bm{\gamma}_{1,c})}}.
\end{align*}
The vector of covariates $\bm{x}_{i}'$ is the same as the one used for the activation patterns in Equation~\eqref{eq:logistic} and $\bm{\gamma}_{1,c}$ is the $1\times m$ vector of coefficients for the period $j=1$ varying with each coverage $c$. 

Once we know which coverages are active in period 1 and which have incurred a payment, we can calculate the corresponding severity per coverage for claim $i$. The actuarial literature contains many examples of the use of Generalized Additive Models for the Location, Scale, and Shape (GAMLSS) for this purpose, thanks to the high level of flexibility that these models provide (see \cite{stasinopoulos2017} for more details on the use of GAMLSS). We use them to predict the expected severity incurred by claim $i$:
\begin{align}
    \expec[Y_{i,1,c} \vert P_{i,1,c}=1] = g^{-1}(\bm{x}_{i}'\bm{\alpha}_{1,c} + \alpha^*_{1,c}1),
\end{align}
where $g(.)$ is the link function and $\bm{\alpha}_{1,c}$ is the $1 \times m$ vector of coefficients for the first development period that depend on the insurance coverages. $\alpha^*_{1,c}$, or more generally for any period $j$, $\alpha^*_{j,c}$, is the specific coefficient attached to the development period in which claim $i$ is at the valuation date. As such, the insurer can apply this severity model in all development periods $j$. The predicted severity will depend on the moment the claim is in its lifetime at the valuation date. 

With the activation patterns, payment patterns, and corresponding severities, we know which coverages claim $i$ activates in development period $j$ and the severity associated with each coverage. 

Note that this model implies that the dependence between the different insurance coverages is accounted for solely by the multinomial logistic regression in Equation~\eqref{eq:logistic}. We further assume the different claims in our dataset to be independent. We only consider dependence between coverages, arising from the activation patterns, while payment patterns and associated severities are all modelled independently.  

\subsubsection{Development periods $j=2, 3, \ldots, j^*$}
\label{sec:j_plus_k}
Knowing what happened in at least one development period, we now describe how the insurer can keep modelling the development of his RBNS claims until they reach development period $j^*$, the period from which he assumes that no more coverage can become active. 

We assume that once an insurance coverage is active, it remains active until settlement of the claim. For claim $i$ and coverage $c$ in development period $j=2, 3, \ldots, j^*$, we then have
\begin{align}
       (A_{i,j,c} \vert A_{i,j-1,c} = 1) = 1, \hspace{0.5cm} 1< j \leq j^*.
\end{align}
As such, we now depict the set of possible patterns for $\bm{A}_{i,j}$ as $\mathcal{V}_j \subset \mathcal{V}_0$, containing all the possible realisations $\bm{v}$. The subset $\mathcal{V}_j$, the possible realisations that it contains, and the number $V_j$ of these possible realisations depend on the activation pattern in development period $j-1$, $\bm{A}_{i,j-1}$.  Knowing this, we use a multinomial logit regression again and write
\begin{align*}
            \prob[&\bm{A}_{i,j} = \bm{v} \vert \bm{A}_{i,j-1}]  \\
            &=\begin{cases}
              \frac{\prob[\bm{A}_{i,j} = \bm{v}]}{\sum_{\bm{v}\in \mathcal{V}_j} \prob[\bm{A}_{i,j} = \bm{v}]}, 
             &\hspace{0.3cm}  \text{if} \hspace{0.2cm}  \bm{v}\in \mathcal{V}_j \\
              0, & \hspace{0.3cm} \text{otherwise,} \numberthis \label{eqn}
            \end{cases}
\end{align*}
where we further assume a Markovian property for the activation pattern vectors and build them using only the information available in the previous period.

Moreover, we assume the Markovian property for the payment patterns, and we have
\begin{align}
   ( P_{i,j,c} \vert A_{i,j,c} = 1)\sim \text{Bernoulli}\big(\pi_{j,c}(\bm{x}_{i}, \bm{\gamma}_{j,c})\big),
\end{align}
where
\begin{align*}
    \pi_{j,c}(\bm{x}_{i}, \bm{\gamma}_{j,c}) = \frac{\exp{(\bm{x}_{i}'\bm{\gamma}_{j,c}})}{1+ \exp{(\bm{x}_{i}' \bm{\gamma}_{j,c})}},
\end{align*}
with the parameter vectors $\bm{\gamma}_{j,c}$ depending once again on the specific insurance coverage $c$ and the vector of risk factors $\bm{x'}_{i}$ is the same $1\times m$ vector of covariates as before.

Knowing which coverages are active in period $j$ and which incurred a payment, we can finally determine the severity of these payments using a similar model to the one used in the first development period.
\begin{align}
\label{eq:GAMLSS}
    \expec[Y_{i,j,c} \vert P_{i,j,c}=1] = g^{-1}(\bm{x}_{i}'\bm{\alpha}_{j,c} + \alpha^*_{j,c}j).
\end{align}

\subsubsection{Development period $j> j^*$}
\label{sec:j_star_plus_k}
\begin{figure}[h!]
\centering
\includegraphics[width=0.8\textwidth]{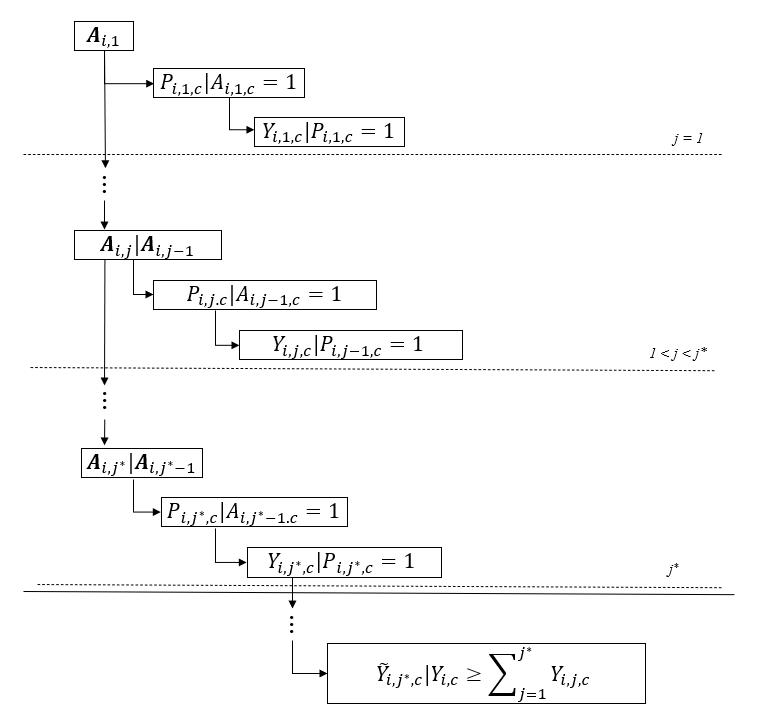}
\caption{Model illustration.
}
\label{fig:Fig5}
\end{figure}

For a given claim $i$ in development period $j > j^*$, the insurer should not use again the model described in Section \ref{sec:j_plus_k} to estimate the activation and payment patterns for the new period since we assume that these will not change anymore after development period $j^*$. Instead, he can directly model the claim settlement by calculating its remaining severity per coverage, $\Tilde{Y}_{i,j^*,c}$, using again GAMLSS as described in Equation \ref{eq:GAMLSS}. $\Tilde{Y}_{i,j^*,c} > 0$ encapsulates the remaining amounts for coverage $c$ at later dates for claims with longer lifetimes and whose activation pattern is not expected to change anymore. 
\newline \newline Figure \ref{fig:Fig5} illustrates the full model described in Section \ref{sec:model}, starting in development period $j=1$ until period $j>j^*$.

\subsection{Simulation routine}
\label{sec:routine}
To properly consider all open claims in a given dataset, we create a simulation routine that we can tailor to the specific development stage of different types of claims. Even though we will solely focus on the RBNS claims for our case study, we present the routine for both IBNR and RBNS claims, including those with longer development times for which we assume that they will not activate any additional coverages.
\paragraph{IBNR claims} 
Let $i$ denote an IBNR claim such that $i \in \mathcal{I}_{\text{IBNR}}$, where the insurer already obtained the set $\mathcal{I}_{IBNR}$ using one of the claims occurrence processes available in the literature. For ease of notation, we drop the subscript $i$ in the description of the routine.
\begin{enumerate}
    \item For development period $j=1$:
    \begin{enumerate}
        \item Simulate the first activation pattern $\bm{A}_{1}$, using the parameters estimated for the multinomial logit model in Section \ref{sec:j}. 
        \item For the inactive coverages, i.e. $A_{1,c} = 0$, then the corresponding $P_{1,c}$ is not defined. For the active coverages only, simulate the payment patterns using the Bernoulli models with parameters $\hat{\pi}_{1,c}$.
        \item If $(P_{1,c} \vert A_{1,c} = 1)=0$, then the corresponding severity for the IBNR claim related to that coverage is not defined (we will set it as zero in the simulations). If $(P_{1,c}\vert A_{1,c} = 1)=1$ from the previous step, we simulate the severity from the distribution selected for the specific coverage.
    \end{enumerate}
    \item For development periods $1 < j \leq j^*$:
    \begin{enumerate}
        \item Knowing the activation pattern $\bm{A}_{j-1}$ for the previous development period, only keep the activation patterns that are now still possible. Re-normalize the probabilities for the remaining activation patterns and simulate them using the multinomial logit model from Section \ref{sec:j_plus_k} to obtain the vector $\bm{A}_{j} \vert \bm{A}_{j-1}$.
        \item If $A_{j,c} = 0$,  $P_{j,c}$ is not defined. For the active coverages only, simulate the payment patterns using the Bernoulli models with parameters $\hat{\pi}_{j,c}$.
        \item If the payment indicator $(P_{j,c} \vert A_{j,c}=1)=0$, the severity for the corresponding coverage is not defined. If $(P_{j,c} \vert A_{j,c}=1)=1$ from the previous step, simulate the severity from the distribution selected for the specific coverage.
        \item Repeat steps (a)-(c) until reaching development period $j^*$. After $j^*$, simulate the remaining severity per coverage, $\Tilde{Y}_{j^*,c}$. 
        The total severity simulated for this IBNR claim is given by$Y^{\text{IBNR}} =  \sum_{j=1}^{j^*} \sum_{c=1}^{C} Y_{j,c} + \sum_{c=1}^{C} \Tilde{Y}_{j^*,c}$. 
    \end{enumerate}
\end{enumerate}

\paragraph{RBNS claims} 
Let $i \in \mathcal{I}_{\text{RBNS}}$. At the valuation date, these claims are already in development period $j>1$. We already know the activation patterns, payment patterns, and payments for development periods $1, \ldots, j-1$. The simulation routine for these claims is then exactly the same as that described for the IBNR claims for development periods $1<j\leq j^*$.

\paragraph{RBNS claims with longer development times}
\normalsize
To remain thorough in our analysis, we must also consider the longer RBNS claims we previously mentioned in Section \ref{sec:j_star_plus_k}. After a certain development period $j^*$, we assume that the activation pattern for claim $i \in \mathcal{I_{\text{RBNS}}}$, $\bm{A}_{i,j^*}$ remains unchanged. If claim $i$ is still open in development period $j^*+k$ for $k \geq 1$, we simulate the remaining severity starting from period $j^*+k$ using the severity model previously described that allows the account for the specific development period in which claim $i$ is at the valuation date. We add the condition that the final severity per coverage for that claim must be at least equal to the sum of the payments observed up to period $j^*+k$, i.e., $(\Tilde{Y}_{i,j^*+k,c} \vert Y_{i,c} \geq \sum_{j=1}^{j^*+k-1} Y_{i,j,c})$, $c=1,2$. The total severity simulated for this RBNS claim is $Y^{\text{RBNS}}_{i} =  \sum_{j=1}^{j^*+k-1} \sum_{c=1}^{C} Y_{j,c} + \sum_{c=1}^{C} \Tilde{Y}_{j^*+k,c}$.

\section{Numerical Application}
\label{sec:application}
Now that we have fully described the statistical model, we dedicate this section to its application to our dataset. We first provide some insights into how we estimate the different model parameters before presenting the results obtained by simulations for the total reserves. We finally compare these reserves to those obtained using two other models that we describe in detail below.

\normalsize
\subsection{Estimation}
\label{sec:estimation}
We apply our model to the data introduced in Section~\ref{sec:data}. We present the estimation results performed for the three components of the model: the activation patterns, payment patterns, and payment severities. 


From our observations in Section~\ref{sec:data}, we assume that the activation patterns in our dataset do not change anymore after development period $j^* = 4$. Considering the information provided in Table~\ref{tab:actidelays}, where we see that almost all claims activate the different insurance coverages no later than four periods after their reporting date, this seems to be a valid assumption.

Appendix \ref{ap:data_sep} illustrates how we separate our dataset into a training and a valuation set, choosing the $1^{\text{st}}$ of January 2019 as valuation date of the reserves. 

\begin{figure}
 \centering
    \includegraphics[width=.8\textwidth]{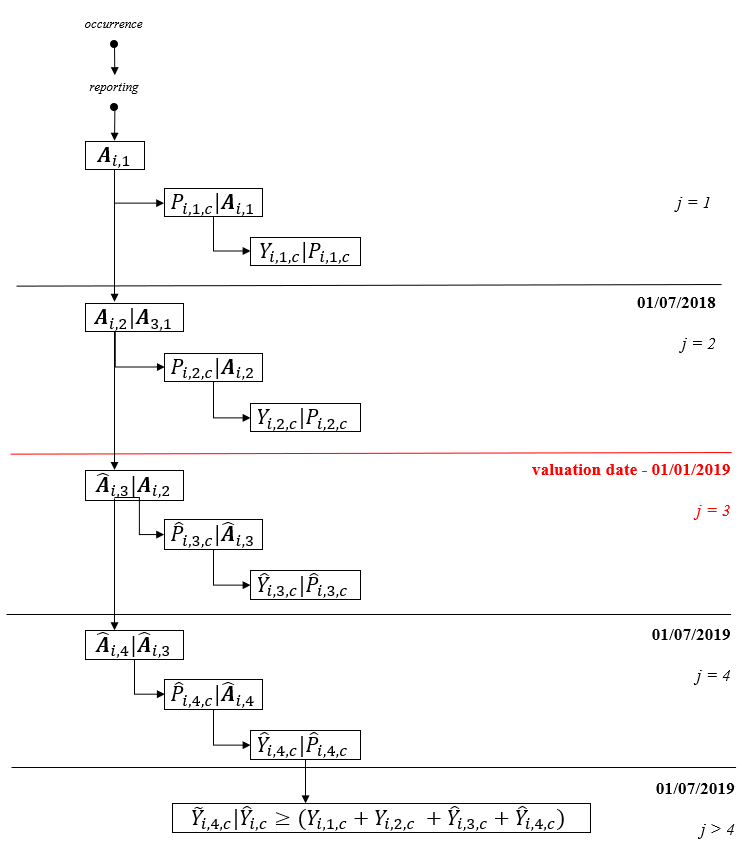}
\caption{Illustration of how the model handles a specific claim from the dataset, taking the $1^{\text{st}}$ of January 2019 as valuation date.}
\label{fig:Fig6}
\end{figure}

Since we have $j^*=4$ and only consider RBNS claims, we use the model starting from Section \ref{sec:j_plus_k} for claims in development periods $j=2$ to $j=4$ and from Section \ref{sec:j_star_plus_k} for claims in development periods $j>4$. Before going further with the model fitting and reserves simulations, we illustrate in Example 2 and Figure \ref{fig:Fig6} how the model handles a given claim from our dataset. \\~\\
\noindent
\small \textit{Example 2. We consider claim $i$, illustrated in Figure~\ref{fig:Fig6}, for which we want to estimate the reserves on January $1^{\text{st}}$, 2019. Claim $i$ was reported in the period ranging from January $1^{\text{st}}$ 2018 to June $30^{\text{th}}$ 2018. This is its first development period, $j=1$. At the valuation date, we observed the activation patterns, payment patterns, and payment severities for the first and second development periods, as claim $i$ enters development period $j=3$. We now have to predict the activation patterns, payment patterns, and payment severities for development periods $j=3$ and $j=4$. Since we assume that the activation patterns do not change anymore after the $4^{\text{th}}$ development period, we can then predict the total remaining severity as of development period $j>4$.}

\normalsize

\subsubsection{Activation patterns}
\label{sec:estimation_A}
Using the same notation as in Section \ref{sec:Model}, we have $C=4$ insurance coverages, leading to $V_0=15$ possible activation patterns. We fit the multinomial logit model for the activation patterns using maximum likelihood estimation with the likelihood function given by
\begin{align}
    \mathcal{L}(\bm{A}_{i,j}) = \prod_{i=1}^{n} \frac{\exp{(\bm{x}_{i}'\bm{\beta}_{j,v})}}{\sum_{k=1}^{V} \exp{(\bm{x}_{i}' \bm{\beta}_{j,k}})}.
\end{align}
In Appendix \ref{ap:multinomial}, we provide estimated parameters for each of the $V_0=15$ possible activation patterns.

We observe that the impact of some risk factors moves away from zero for the $5^{\text{th}}$ pattern, i.e., the one in which a claim simultaneously activates the Bodily Injury and Loss of Use coverages. In particular, due to the legislation regarding automobile insurance in Quebec and Saskatchewan, the probability of observing that specific activation pattern decreases substantially if the claim occurred in one of these provinces. In the province of Quebec, the Bodily Injury coverage does not exist: a person injured in a car accident will benefit from the public automobile insurance plan provided by the Société de l'assurance automobile du Québec (SAAQ) rather than from a private insurance company. In Saskatchewan, the insured must choose between a \textit{no-fault} and a \textit{tort} auto injury insurance coverage offered by the Saskatchewan driver's licensing and vehicle registration. Almost all residents choose the \textit{tort} coverage under which they are insured regardless of whether they are at fault or not. As such, the Bodily Injury coverage rarely appears for claims in that province. Consequently, we see that the parameter estimates for Quebec and Saskatchewan are always under $0$ for all activation patterns, including the Bodily Injury coverage. 

Moreover, the birth year is also a big driver for this particular pattern: the probability of observing it increases for the younger insureds born in the 1990s and after 2000. We commonly observe in automobile insurance data that younger drivers tend to cause more frequent and more severe accidents. Consequently, we can expect more accidents with injuries or even casualties and loss of use in the younger population. Appendix \ref{ap:multinomial} illustrates this well, where the insureds born after 2000 are the only cohort that always has an increasing impact on the probability of observing each activation pattern, even though they only represent $1.52\%$ of all the insureds.

\subsubsection{Payment patterns}
We fit twelve Bernoulli regressions for the payment patterns: one for each of the four coverages in development periods $j = 2$, $j = 3$, and $j = 4$ and use again maximum likelihood optimisation to obtain parameter estimates for each of the models. Table \ref{tab:paymentpatterns} displays the results of these estimations. 
\begin{table}[h]
\begin{center}
\caption{Fitted average probabilities of observing a payment}
\label{tab:paymentpatterns}
\begin{tabular}{@{}l r r r@{}}
\toprule
  Probability & \multicolumn{1}{c}{$j=2$} & \multicolumn{1}{c}{$j = 3$} & \multicolumn{1}{c}{$j = 4$}\\
\midrule
 $\pi_{j,\text{AB}}$ &  0.2367 & 0.2194 & 0.1267\\
 $\pi_{j,\text{BI}}$ &  0.1426 & 0.1220 & 0.0695\\
 $\pi_{j,\text{VD}}$ &  0.6391 & 0.1030 & 0.0311\\
 $\pi_{j,\text{LoU}}$ &  0.4228 & 0.0368 & 0.0087 \\
 \bottomrule
\end{tabular}
\end{center}
\end{table}

\subsubsection{Payment severities}
\begin{figure}[h!]

\centering
\includegraphics[width=.9\textwidth]{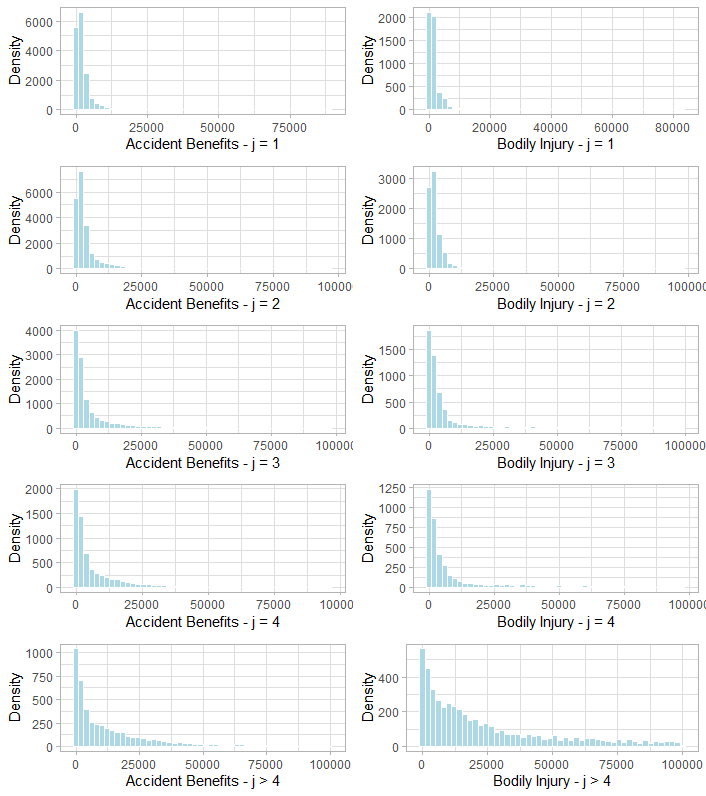}
\caption{Histograms of the observed payments for the Accident Benefits and Bodily Injury coverages}
\label{fig:Fig7}
\end{figure}

From Table \ref{tab:infocov}, we note the importance of considering heavy-tailed distributions to model the severity of payments, as is usually done in the actuarial literature. Among others, Frees et al. (2009) \cite{frees2009} opt for the Generalized Beta of the second kind distribution to accommodate the long-tailed nature of claims. Figure \ref{fig:Fig7} shows the histogram of the payments made in the different development periods considered in the model for the Accident Benefits and Bodily Injury coverages. Note that for both of them, we cut out most of the tail of the histograms in the right part of the graph to ease the readability. 

For each coverage, we consider five commonly used distributions: the Log-Normal, Gamma, Pareto, Generalized Beta of the second kind, and Weibull distributions. As explained in Section \ref{sec:Model}, we fit each of these models for all possible development periods, adding a covariate that specifies the development period in which a given claim is at the valuation date. Appendix \ref{ap:aic_bic} presents the Akaike Information Criterion (AIC) and Bayesian Information Criterion (BIC) for all models and all four insurance coverages. We conclude that the preferred distribution for the Accident Benefits, Vehicle Damage, and Loss of Use coverages is the Generalized Beta of the Second Kind. We select the Weibull distribution for the Bodily Injury coverage.

\subsection{Predictive distributions and model comparisons}
We focus in this section on presenting the results obtained for the estimation of the reserves when we apply the model described in Section \ref{sec:Model} to the dataset described in Section \ref{sec:data}, choosing the $1^{\text{st}}$ of January 2019 as the valuation date. We present the predictive distribution of the reserves for the portfolio as a whole and the different insurance coverages. We compare these results to the observed reserve amount and the one obtained from fitting some more classical and commonly used reserving models.

For all models we consider in this section, we perform at least 5,000 simulations using the simulation routine described in Section \ref{sec:routine}. As shown in Appendix \ref{ap:stability}, this large number brings stability to our results. We first present the estimations from the activation patterns model before comparing them to other reserving models. 

\begin{figure}[h!]
\centering
\includegraphics[width=0.6\textwidth]{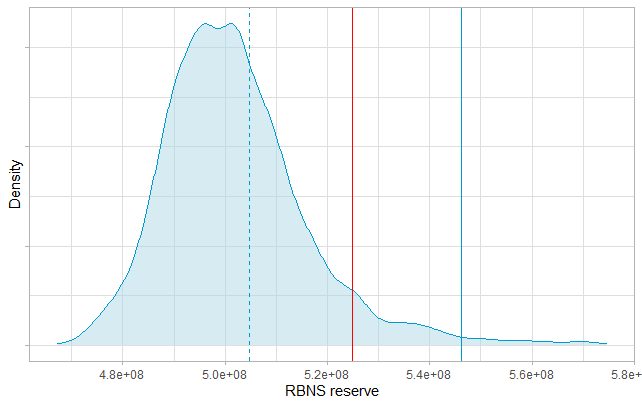}
\caption{Simulated RBNS reserves for the full portfolio as of the 1st of January 2019. The red line, dashed blue line and continuous blue lines depict, respectively, the observed reserve amount, the average and the $0.95$ quantile of the simulations}
\label{fig:Fig8}
\end{figure}

Figure \ref{fig:Fig8} displays the predictive distribution of the total RBNS reserve of our portfolio. In this graph and all those that will follow, the red line marks the observed reserve, the blue dotted line shows the average value of the predictions, and the continuous blue line depicts the $0.95$ quantile of the distribution. 

For these claims, the observed reserve amount is 524.91M~CAD. Note that the true amounts that we provide henceforth are minimum amounts because more than $1\%$ of the claims are still open in the portfolio at the valuation date. The final severity for these claims will probably be greater than the one observed now once they become settled.  

\begin{figure}[h!]
\centering
\includegraphics[width=1\textwidth]{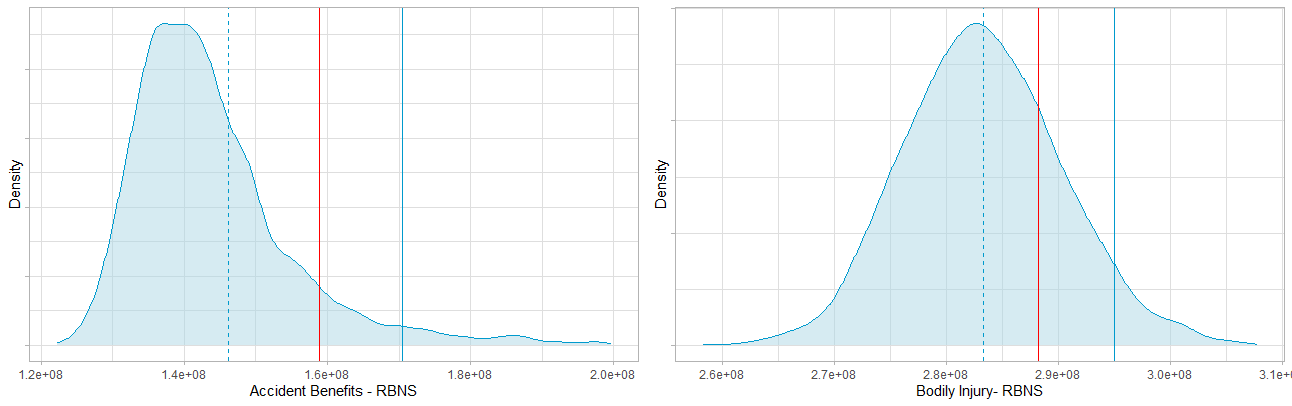}
\caption{Simulated RBNS reserves for the Accident Benefits (left) and Bodily Injury (right) coverages. The red lines, dashed blue lines and continuous blue lines depict, respectively, the observed reserve amounts, the average and the $0.95$ quantile of the simulations}
\label{fig:Fig9}
\end{figure}

Figure \ref{fig:Fig9} presents the predictive distributions obtained for the reserves of the Accident Benefits and Bodily Injury insurance coverages.
We provide further detail on these results in Table \ref{tab:apm_results}. In particular, we display the predicted reserves for the four coverages and the whole portfolio using six-month and one-year periods, the latter being more commonly used by academics and practitioners alike. In the remainder of this section, we will only use the results obtained with the six months periods for the activation patterns model and other models used for comparison. Since financial legislation around the world typically requires insurers to set aside an amount based on a high quantile of the predictive distribution of their reserves, we provide the $95\%$ Value-at-Risk (VaR) in addition to the mean for each of the four coverages and the portfolio as a whole. 
\begin{table}[h]
\begin{center}
\caption{Simulated RBNS reserves (in M~CAD) using the activation patterns with development periods of 6 months (columns 3 and 4) and 1 year (columns 5 and 6).}
\label{tab:apm_results}
\begin{tabular}{@{}l r r r r r r@{}}
\toprule
\multirow{2}{*}{Coverage} & \multirow{2}{*}{True reserves} & \multicolumn{2}{c}{Act. Patterns (6 mo.)} & \multicolumn{2}{c}{Act. Patterns (1 y.)}  \\
\cline{3-6}
 &  & \multicolumn{1}{c}{Mean} & \multicolumn{1}{c}{$\text{VaR}_{0.95}$}   & \multicolumn{1}{c}{Mean} & \multicolumn{1}{c}{$\text{VaR}_{0.95}$}  \\
\toprule
Accident Benefits & $>$158.90 & 146.61 & 170.46 & 162.15 & 196.64\\
Bodily Injury & $>$288.17 & 283.27 & 295.02 & 300.02 & 312.58 \\
Vehicle Damage & $>$73.63 & 70.77 & 76.43 & 70.80 & 72.95 \\
Loss of Use & $>$4.20 & 4.17 & 4.24 & 3.41 & 4.01 \\
\midrule
Global reserve & $>$524.91 & 504.82 & 546.15 & 536.86 & 574.58 \\
\bottomrule
\end{tabular}
\end{center}
\end{table}
\newline 
For each coverage and the portfolio, the average predictions underestimate the observed reserves, while the predictions obtained for the $95\%$ VaR are all greater than the observed amounts. This is particularly the case for the Accident Benefits and Bodily Injury coverages where the VaR amounts are, respectively, 170.46M CAD and 295.02M CAD, while the observed reserves are equal to 158.90M CAD and 288.17M CAD, respectively. For the Loss of Use coverage that represents less than $1\%$ of the total reserve, the quantile of the predictions is close to the observed amount.

Table \ref{tab:apm_results} illustrates that for a short-tailed claims dataset such as ours, using 6-month rather than one-year development periods leads to more accurate predictions, both for the portfolio as a whole and for each separate coverage.
\newline \newline
Next, we pursue the analysis of our results by comparing them to those obtained using a classic aggregate model and a replica of the activation patterns model that does not take dependence into account.

\subsubsection{Activation patterns model vs aggregate models}
\label{sec:APMvsODP}
Table \ref{tab:apm_vs_ODP} compares the reserve estimates obtained with the proposed model based on the activation patterns to those of the classical Overdispersed Poisson (ODP) Chain Ladder and to the observed reserves, both for the portfolio and the four insurance coverages taken separately. Since we only focus our analysis on RBNS claims and for the sake of comparison with the other models, we use reporting years instead of occurrence years when building the claims triangles. We further assume that all underlying assumptions of the ODP model are verified but do not validate them here since this is a standard model in claims reserving and our aim is to focus on a comparison with the activation patterns model.

Note that in Table \ref{tab:apm_vs_ODP}, the global reserve for the ODP model is not equal to the sum of the reserves per coverage. It is due to building the claims development triangles and applying the model to each coverage taken separately, as well as to the portfolio, rather than just taking the sum of the reserves obtained per coverage to obtain the total reserve. Before taking a closer look at the results obtained with the Overdispersed Poisson Chain Ladder, we remind the reader that we only have a limited number of calendar years observed in our dataset with a claim history spanning from 2015 to mid-2021. Even if we consider development periods of six months rather than the more traditional one-year periods, our claims development triangles are still based on a small number of claims. We must be careful with the results obtained with an aggregate reserving method, particularly when we apply it to coverages such as the Accident Benefits and Bodily Injury, because these methods typically require large data sets for a more in-depth analysis. To compensate for this lack of data and for the longer tailed nature of these two coverages, we use tail factors obtained from the simulations performed with the activation patterns model. This allows for a more suitable comparison with the results of the other two models discussed in this section. 

At the portfolio level, the aggregate model leads to predictions for the total reserves slightly above the observed amount, with a $95\%$ VaR equal to 529.71M CAD. If instead of applying the model to the portfolio as is usually done, we implement it to each coverage separately and then sum up the predictions obtained, we reach an average portfolio reserve and a $95\%$ VaR equal to, respectively, 449.28M CAS and 469.58M CAD. Neither of these values are sufficient to cover the observed reserve. This highlights the disadvantages of using an aggregate model such as the ODP Chain Ladder to a more ganular level of the dataset.

The average and quantile predictions underestimate the observed reserves for all coverages except the Loss of Use coverage where the $95\%$ VaR is slightly above the true amount of 4.20M CAD. The aggregate model provides overall lower predicted values than the activation patterns model and mostly falls short of the observed amounts.
\begin{table}[h!]
\begin{center}
\caption{Comparison between the activation patterns model and the Overdispersed Poisson Chain Ladder (in M~CAD)}
\label{tab:apm_vs_ODP}
\begin{tabular}{@{}l r r r r r @{}}
\toprule
\multirow{2}{*}{Coverage} &  \multirow{2}{*}{True reserve} &  \multicolumn{2}{c}{Activation patterns} & \multicolumn{2}{c}{ODP}  \\
\cline{3-6}
 &   & \multicolumn{1}{c}{Mean} & \multicolumn{1}{c}{$\text{VaR}_{0.95}$} & \multicolumn{1}{c}{Mean} & \multicolumn{1}{c}{$\text{VaR}_{0.95}$} \\
\midrule
Accident Benefits  & $>$158.90 & 149.61 & 170.46 & 148.73 & 153.41 \\
Bodily Injury & $>$288.17 & 283.27 & 295.02 & 241.43 & 253.23  \\
Vehicle Damage & $>$73.63 & 70.77 & 76.43 & 55.17 & 58.61  \\
Loss of Use  & $>$4.20 & 4.17 & 4.24 & 3.95 & 4.33 \\
\midrule
Global reserve & $>$524.91 & 504.82 & 546.15 & 515.62 & 529.71 \\
\bottomrule
\end{tabular}
\end{center}
\end{table}

\subsubsection{Activation patterns model vs independence model}
\label{sec:APMvsIND}
We perform independence tests on our data to strengthen our argument in favour of a model that considers the possible dependence between coverages. First, we test for independence between each pair of insurance coverages using the chi-squared goodness of fit test. Each of these two-way interaction tests rejects the hypothesis of independence. We also perform likelihood ratio tests comparing models with no interaction, two, three, and four-way interactions. Once again, the tests show the presence of dependence between the coverages and reject the simpler models in favour of the more complex ones in each scenario. 

We build an independence model to assess the impact of modelling this dependence on the reserve estimates. We reproduce the activation patterns model but rather than modelling the activation patterns in development periods $j=2$ to $j=4$ with the multinomial logit models described in Sections \ref{sec:j} and \ref{sec:j_plus_k}, we model the activation of the coverages using separate and independent Bernoulli regressions for development periods $j=2$ to $j= 4$. We fit twelve Bernoulli regressions: one for each of the four insurance coverages and the three development periods considered for RBNS claims. The rest of the model, i.e., the estimation of the payment patterns, payment severities, and all the assumptions linked to them, remain the same. We present the results obtained with this model in Table \ref{tab:apm_vs_ind} and compare them to the observed reserves and the estimates from our activation patterns model. 

Note that since we use the same severity model in the activation patterns and independence models, the average predictions of both models are very close to each other for all insurance coverages and the portfolio. The differences arise in the tails of the predictive distributions where VaRs obtained with the independence model are not always higher than observed amounts, contrary to the ones obtained with the activation patterns model. It is the case for the Loss of Use and, more importantly, Accident Benefits coverages. With an average predicted amount and $95\%$ VaR of, respectively, 507.8M CAD and 533.90M CAD, the independence model provides overall lower estimates for the portfolio Value-at-Risk than the activation patterns model compared to the observed reserve. We also note that predictive distributions for the Accident Benefits coverage, the Bodily Injury coverage, and the whole portfolio are less heavy-tailed than those from the activation patterns model. 
\begin{table}[h!]
\begin{center}
\caption{Comparison between the activation patterns model and the independence model (in M~CAD)}
\label{tab:apm_vs_ind}
\begin{tabular}{@{}l r r r r r@{}}
\toprule
\multirow{2}{*}{Coverage} &  \multirow{2}{*}{True reserve} &  \multicolumn{2}{c}{Activation patterns} & \multicolumn{2}{c}{Independence}  \\
\cline{3-6}
 &   & \multicolumn{1}{c}{Mean} & \multicolumn{1}{c}{$\text{VaR}_{0.95}$} & \multicolumn{1}{c}{Mean} & \multicolumn{1}{c}{$\text{VaR}_{0.95}$} \\
\midrule
Accident Benefits  & $>$158.90 & 149.61 & 170.46 & 151.72 & 158.23 \\
Bodily Injury & $>$288.17 & 283.27 & 295.02 & 283.03  &  292.95  \\
Vehicle Damage & $>$73.63 & 70.77 & 76.43  & 68.99 & 78.59 \\
Loss of Use  & $>$4.20 & 4.17 & 4.24 & 4.06 & 4.13 \\
\midrule
Global reserve & $>$524.91 & 504.82 & 546.15  & 507.8 & 533.90 \\
\bottomrule
\end{tabular}
\end{center}
\end{table}

\subsubsection{Results summary}   
Table \ref{tab:summary_models} summarises the results discussed in Sections \ref{sec:APMvsODP} and \ref{sec:APMvsIND}.

The predictions from a reserving model should be sufficient to cover all future liabilities for incurred but not reported claims and, as in this paper, reported but not settled claims at the chosen valuation date. 

With average predictions for the total reserves of 504.82M CAD, 515.62M CAD, and 507.8M CAD for, respectively, the activation patterns, ODP Chain Ladder, and independence models, none of these models provide estimates that are sufficient to cover the observed global portfolio reserve of 524.91M CAD. However, all three models allow us to obtain $95\%$ VaRs that seem large enough to cover this amount. When applied to the portfolio, the ODP Chain Ladder provides the lowest global VaR with a value that is just above the observed reserve. The independence model predicts a $95\%$ VaR of 533.90M CAD, and the activation patterns model finally presents the highest value for the quantile with a reserve amount of 546.15M CAD. 

When looking at the reserves per coverage, the activation patterns model is the only one that provides Values-at-Risk higher than the observed reserves in every case. The quantiles of the predictive distributions in the case of the independence model fall short of the observed amounts for the Accident Benefits and Loss of Use coverage, and quantiles of the aggregate model underestimate the reserve for all coverages except Vehicle Damage.

\begin{table}[h!]
\begin{center}
\caption{Model comparison - summary (in M~CAD)}
\label{tab:summary_models}
\begin{tabular}{@{}l r l r r r r @{}}
\toprule
Coverage  & \multicolumn{1}{c}{Observed}  &  \multicolumn{1}{c}{Simul.} & \multicolumn{1}{c}{Act. pat.} & \multicolumn{1}{c}{ODP}  & \multicolumn{1}{c}{Ind.}  \\
\midrule
\multirow{2}{*}{Accident Benefits}& \multirow{2}{*}{$>$158.90}  & Mean & 149.61 & 148.73 & 151.72  \\
&  &$\text{VaR}_{0.95}$ &  170.46 & 153.41 & 158.23 \\
\multirow{2}{*}{Bodily Injury} & \multirow{2}{*}{$>$288.17} & Mean  & 283.27  & 241.43  & 283.03   \\
&  &$\text{VaR}_{0.95}$ &  295.02  & 253.23  & 292.95\\
\multirow{2}{*}{Vehicle Damage} & \multirow{2}{*}{$>$73.63} & Mean & 70.77 & 55.17  & 68.99 \\
& & $\text{VaR}_{0.95}$& 76.43 & 58.61 & 78.59  \\
\multirow{2}{*}{Loss of Use}  & \multirow{2}{*}{$>$4.20} & Mean  & 4.17 & 3.95  & 4.06 & - \\
& & $\text{VaR}_{0.95}$& 4.24  & 4.33  & 4.13 \\
\midrule
\multirow{2}{*}{Global reserve} & \multirow{2}{*}{$>$524.91} & Mean & 504.82 & 515.62 & 507.80\\
& & $\text{VaR}_{0.95}$ & 546.15  & 529.71  & 533.90 \\
\bottomrule
\end{tabular}
\end{center}
\end{table}

\section{Conclusion}
\label{sec:conclusion}
In this paper, we analyze a Canadian automobile dataset in which each claim can impact one or more of four insurance coverages provided by the company, namely Accident Benefits, Bodily Injury, Vehicle Damage, and Loss of Use. We seek to predict the claims reserve for this portfolio while considering the underlying dependence between these coverages. More specifically, we analyze how a single claim can simultaneously activate multiple coverages and the potential impact this can have on the final amount of the reserve.

To do this, we build an individual model based on activation patterns for the insurance coverages to estimate claims reserves. Based on the predictions of the so-called activation patterns of the four coverages in a given development period $j$, we then predict whether an active coverage will incur a payment and its corresponding amount. We capture the dependence between the coverages in the activation patterns with multinomial logit regressions. Even though we tailor our model to our specific study, we present it in the most general way possible such that one can easily extend it to other applications. For instance, the number of insurance coverages, the number of development periods after which the activation patterns remain stable, the length of the development period, or the severity distributions are all features of the model that one can adapt to other situations. Further developments of this model could include the addition of the age of the claim as a risk factor to account for time dependence. We could also drop the assumption of independence between the claims and seek to model it in a similar way as the dependence between coverages.

We then compare the results obtained for our dataset with those from two additional models: the classical Overdispersed Poisson Chain Ladder and a replica of the activation patterns model that does not consider dependence. Specifically, this replica, named the \textit{independence model} in the paper, predicts the activation patterns of the four coverages using four independent Bernoulli regressions per period rather than a multinomial logit model. It guarantees that the activation of a coverage does not impact the activation of the other coverages in a given period.

We observe in the results that, while all three models provide predictions for the $95\%$ portfolio VaRs of the global reserve larger than the observed amount, only the predicted values from the activation patterns model appear to be above the observed amounts as well for each coverage taken separately. While the other methods also allow us to obtain large enough predictions for the total reserves, they seem less efficient at the coverage level. 

With the activation patterns model, we aim to better highlight and understand the underlying dynamics of the claims portfolio at hand through the activation of the different insurance coverages. It is a good illustration of the challenge faced nowadays by many practitioners in the insurance industry that are required, either by regulators or their own internal processes, to predict their claims reserves with increasing levels of granularity. 


\clearpage

\appendix
\section*{\Large Appendix}

\section{Observed activation patterns}
\label{ap:observed_acti}
Table \ref{tab:actipatterns} presents the activation patterns $\bm{A}_{i,1}$ observed in the dataset for the first development period of the claims. The value $1$ stands for the activation of the coverage. $85.67\%$ of all claims activate the Vehicle Damage coverage alone or simultaneously with the Loss of Use coverage. The Accident Benefits coverage is most often triggered with the Vehicle Damage coverage or on its own. We also observe that around $1\%$ of all claims simultaneously activate all four coverages upon their reporting. 
\begin{table}[h]
\begin{center}
\caption{Frequency of the activation patterns observed in the first development period}
\label{tab:actipatterns}
\begin{tabular}{@{}r r r r r @{}}
\toprule
  \multicolumn{1}{c}{Accident Benefits} & \multicolumn{1}{c}{Bodily Injury} & \multicolumn{1}{c}{Vehicle Damage} & \multicolumn{1}{c}{Loss of Use}  & \multicolumn{1}{c}{$\%$ of claims} \\
 \midrule
 0  & 0 & 1 & 0 & 42.95 \\
 0  & 0 & 1 & 1 & 42.72 \\
 1  & 0 & 1 & 0 & 4.98 \\
 0  & 1 & 1 & 1 & 1.99 \\
 1  & 0 & 0 & 0 & 1.44 \\
 0  & 1 & 1 & 0 & 1.24 \\
 1  & 0 & 1 & 0 & 1.07 \\
 0  & 0 & 0 & 1 & 1.04 \\
 1  & 1 & 1 & 1 & 1.01 \\
 0  & 1 & 0 & 0 & 0.61 \\
 1  & 1 & 1 & 0 & 0.43 \\
 1  & 1 & 0 & 0 & 0.38 \\
 1  & 0 & 0 & 1 & 0.10 \\
 0  & 1 & 0 & 1 & 0.04 \\
 1  & 1 & 0 & 1 & 0.01 \\
 \bottomrule
\end{tabular}
\end{center}
\end{table}

\clearpage
\section{Risk factors}
\label{ap:riskfactors}
\begin{figure}[h!]
    \centering
    \includegraphics[width=1\textwidth]{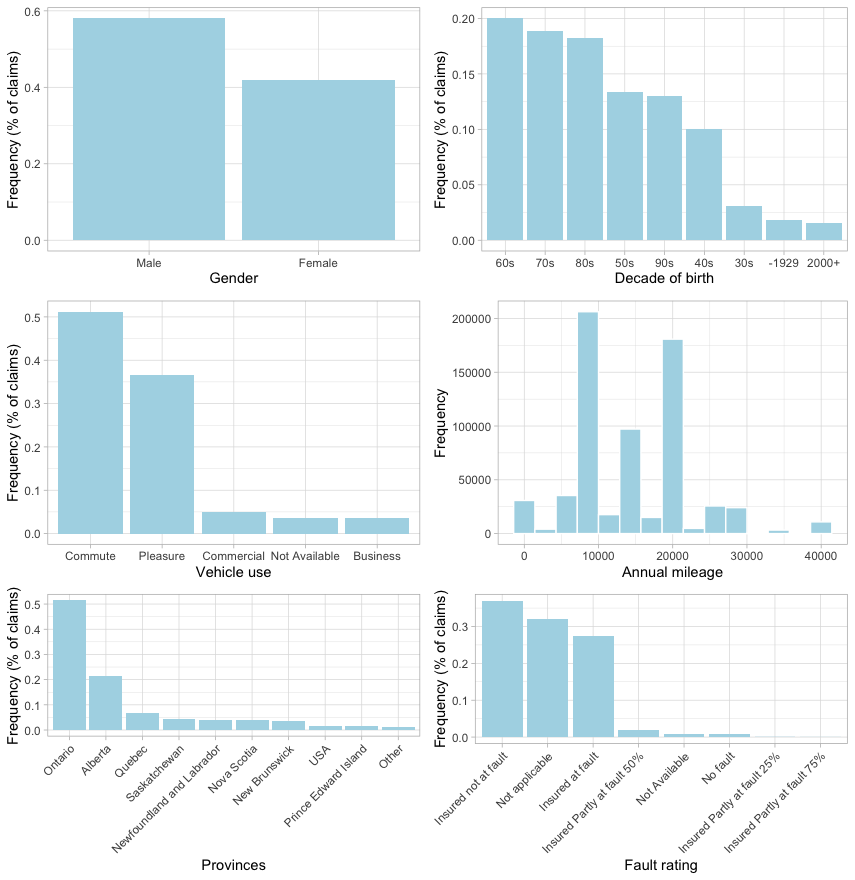}
    \caption{Risk factors}
    \label{fig:Fig10}
\end{figure}

Figure \ref{fig:Fig10} completes Section \ref{sec:riskfactors} by providing further insights on the different risk factors used in the activation patterns model. 
\newline Our dataset includes around $60\%$ of male insureds against $40\%$ of female insureds, more than half born between 1960 and 1989. Approximately half of the insureds use their vehicle for commuting, while an additional $37\%$ use it for pleasure. The remaining $13\%$ of the insureds use their car for commercial or business reasons or did not disclose that information to the insurer. $76\%$ of the insureds drive between 10,000 and 20,000 kilometers per year according to the annual mileage risk factor, which is the only continuous one in our dataset. More than $50\%$ of the claims occur in Ontario, $20\%$ in Alberta, and the remaining $30\%$ in other Canadian provinces or in the United States of America. Finally, the insurance company assesses the insured's level of responsibility in an accident and issues a fault rating. In $37\%$ of the claims, the insurance company considered the insured not to be at fault. The company could not apply the rating for $32\%$ of the claims, and in $27\%$ of the cases, the insured bore the entire fault. For the remaining $4\%$ of the claims, there was either no fault at all or the fault was divided between the insured and any other parties involved in the claim.

\newpage
\section{Training and valuation sets}
\label{ap:data_sep}
\begin{figure}[h!]
    \centering
    \includegraphics[width=.9\textwidth]{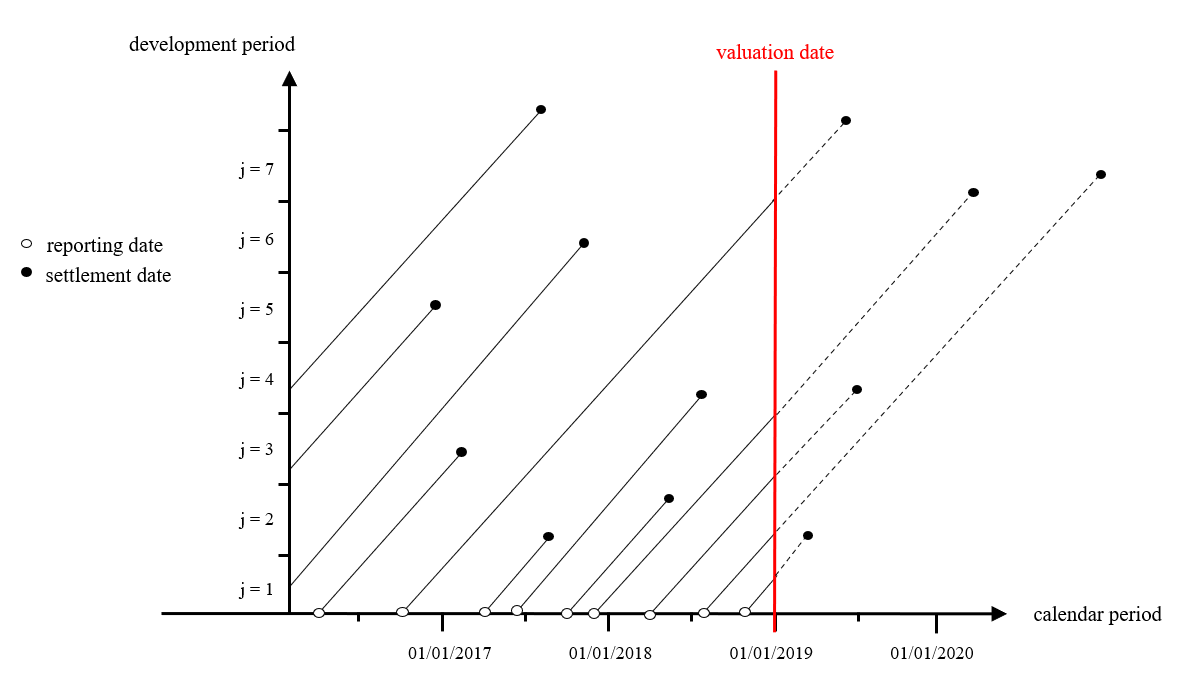}
    \caption{Illustration of the separation of the dataset into the training and valuation sets.}
    \label{fig:Fig11}
\end{figure}

To fit the activation patterns model, we split the data into a training set for estimation, and a valuation set for simulation, as illustrated in Figure~\ref{fig:Fig11}.  In this Figure, each black continuous (resp. dashed) line symbolizes the observed (resp. future) development of a claim.

The red vertical line in Figure \ref{fig:Fig11} represents the chosen valuation date of the reserves. All the information displayed on the left of this valuation date (continuous lines) is the observed claims data at the valuation date. Since we only focus on RBNS claims in this paper, it contains all the claims that were reported before the valuation date. For those already settled, it includes the complete claim development. For those still open at the valuation date, it contains all the information (i.e., activation of coverages, payments made, and corresponding amounts) observed up to that date. All the information on the right of the red line, represented by the dashed lines, makes up the valuation dataset we will use to predict the reserves. 
Our dataset also contains around $1\%$ of open claims, i.e., claims not yet settled on the $31^{\text{st}}$ of June 2021. We include these claims by taking their last observed payment as the settlement date.

\section{Multinomial logit model}
\label{ap:multinomial}
\begin{sidewaystable}
\begin{center}
\footnotesize
\caption{Parameter estimates for the multinomial logit model}
\label{tab:A1fit}
\scalebox{1}{
\begin{tabularx}{610pt} { l l r r r r r r r r r r r r r r}
\toprule
\multicolumn{2}{c}{Risk factors}  & \multicolumn{1}{c}{2} & \multicolumn{1}{c}{3} & \multicolumn{1}{c}{4} & \multicolumn{1}{c}{5} & \multicolumn{1}{c}{6} & \multicolumn{1}{c}{7} & \multicolumn{1}{c}{8} & \multicolumn{1}{c}{9} & \multicolumn{1}{c}{10} & \multicolumn{1}{c}{11} & \multicolumn{1}{c}{12} & \multicolumn{1}{c}{13} & \multicolumn{1}{c}{14} & \multicolumn{1}{c}{15}  \tabularnewline
\midrule
\multicolumn{2}{l}{Intercept} & 3.45 & 3.44 & -0.26 & -12.78 & 1.31 & 1.93 & 0.30 & -1.04 & 1.14 & 1.67 & 0.28 & -1.68 & 0.56 & 1.69\tabularnewline 
\midrule
\multirow{9}{*}{PROV} & New Brunswick & 0.21 & -0.16 & 1.11 & 0.43 & 0.05 & -0.49 & -0.93 & 0.29 & -0.18 & 0.35 & 0.48 & 1.07 & 0.40 & 0.35 \tabularnewline
 & Newfoundland and Labrador & 0.55 & 0.02 & 0.58 & 0.63 & 0.09 & 0.27 & -0.06 & -0.16 & -0.09 & 0.26 & 0.79 & 0.83 & 0.61 & 0.65 \tabularnewline
 & Nova Scotia & 0.36 & -0.05 & 1.09 & 0.18 & -0.08 & -0.72 & -1.08 & 0.23 & -0.73 & 0.30 & 0.48 & 0.01 & -0.01 & 0.29 \tabularnewline
 & Ontario & -0.14 & -0.12 & 0.93 & 0.25 & -0.35 & -0.32 & -0.50 & -0.15 & -1.26 & -0.34 & 0.15 & 0.41 & -0.65 & -0.31 \tabularnewline
 & Other & 0.57 & -0.14 & 0.61 & -0.62 & -0.06 & -0.08 & 0.25 & 0.10 & 0.34 & -0.28 & 0.76 & 1.87 & 0.71 & 0.29\tabularnewline
 & Prince Edward Island & 0.93 & -0.13 & 0.93 & 0.98 & 0.17 & -0.75 & -0.57 & 0.31 & 0.01 & 0.31 & 0.58 & -3.21 & 1.11 & 0.61 \tabularnewline
 & Quebec & 1.64 & -0.11 & -2.10 & -19.74 & -2.77 & -3.31 & -2.34 & -2.10 & -1.97 & -2.01 & -1.66 & -0.22 & -2.46 & -2.34 \tabularnewline
 & Saskatchewan & 1.53 & -0.40 & -1.21 & -23.02 & -1.04 & -2.84 & -1.05 & -0.78 & -0.41 & -1.27 & -2.73 & -390.59 & -0.60 & -2.37 \tabularnewline
 & USA & -5.52 & -1.50 & -3.83 & 11.02 & -5.16 & -2.24 & -7.22 & 0.32 & -5.19 & -0.99 & -3.53 & 0.10 & -3.45 & -0.86  \tabularnewline
\midrule
\multirow{8}{*}{YOB} & 2000+ & 0.71 & 0.70 & 3.71 & 14.79 & 3.87 & 3.80 & 3.90 & 2.92 & 3.61 & 3.48 & 3.94 & 3.49 & 3.84 & 3.62 \tabularnewline
 & 1930s & -0.34 & -0.21 & -1.13 & 9.95 & -1.09 & -0.84 & -0.96 & -1.10 & -0.72 & -1.0 & -1.47 & -1.95 & -1.07 & -1.04 \tabularnewline
 & 1940s & -0.41 & -0.26 & -0.72 & 10.46 & -0.49 & -0.56 & -1.39 & -1.82 & -1.16 & -1.08 & -1.90 & -2.24 & -1.68  & -1.64 \tabularnewline
 & 1950s & -0.55 & -0.27 & -0.82 & 10.36 & -0.56 & -0.73 & -0.84 & -1.57 & -0.69 & -0.82 & -1.54 & -1.81 & -1.36 & -1.29  \tabularnewline
 & 1960s & -0.55 & -0.29 & -0.58 & 10.44 & -0.33 & -0.44 & -0.81 & -1.31 & -0.82 & -0.88 & -1.79 & -1.36 & -1.57 & -1.60 \tabularnewline
 & 1970s & -0.60 & -0.26 & -0.54 & 10.37 & -0.30 & -0.37 & -0.75 & -1.32 & -0.74 & -0.81 & -1.88 & -1.43 & -1.52 & -1.51 \tabularnewline
 & 1980s & -0.73 & -0.33 & -0.48 & 10.18 & -0.29 & -0.40 & -0.69 & -1.14 & -0.78 & -0.71 & -1.62 & -1.18 & -1.27 & -1.26 \tabularnewline
 & 1990s & -0.44 & -0.30 & -0.30 & 11.04 & -0.08 & -0.12 & 0.15 & -1.20 & -0.07 & -0.48 & -0.50 & -0.50 & -0.39 & -0.63 \tabularnewline
 \midrule
 GENDER & Male & 0.24 & 0.08 & -0.22 & -0.44 & -0.23 & -0.28 & -0.26 & -0.89 & -0.35 & -0.89 & -0.28 & -0.69 & -0.45 & -0.92 \tabularnewline
 \midrule
 \multicolumn{2}{l}{AM} & 0.00 & 0.00 & 0.00 & 0.00 & 0.00 & 0.00 & 0.00 & 0.00 & 0.00 & 0.00 & 0.00 & 0.00 & 0.00 & 0.00 \tabularnewline
 \midrule
\multirow{4}{*}{VU} & Commercial & 0.68 & -0.11 & 0.83 & 0.25 & 0.71 & -0.19 & 0.28 & -0.299 & 0.89 & -0.19 & 0.71 & -0.55 & 0.84 & -0.27 \tabularnewline
 & Commute & -0.13 & -0.00 & -0.08 & -0.20 & -0.05 & -0.14 & 0.03 & -0.11 & 0.15 & 0.09 & 0.07 & -0.34 &  0.17 & -0.02\tabularnewline
 & Not Available & 2.37 & 0.13 & 2.73 & -0.07 & 2.59 & 0.28 & 2.25 & -1.88 & 1.44 & -1.13 & 0.11 & -3.26 & 1.28 & -1.50 \tabularnewline
 & Pleasure & 0.15 & -0.03 & -0.00 & -0.15 & 0.05 & -0.21 & 0.25 & -0.22 & 0.57 & 0.04 & 0.34 & -0.46 & 0.57 & -0.06\tabularnewline
 \midrule
\multirow{7}{*}{FR} & Insured not at fault & -0.94 & -0.06 & -2.53 & -2.32 & -3.38 & -2.70 & 0.55 & 0.84 & -0.23 & 0.72 & -1.28 & -1.39 & -1.82 & -1.35 \tabularnewline
 & Insured Partly at fault $25\%$ & 1.10 & 1.37 & 2.96 & -12.13 & 1.62 & 2.13 & 0.89 & -9.41 & 2.23 & 2.10 & 3.51 & -2.41 & 3.43 & 2.98 \tabularnewline
 & Insured Partly at fault $50\%$& 0.34 & 0.56 & -0.01 & -0.42 & -0.12 & -0.01 & -0.92 & -0.38 & 0.18 & 0.71 & -0.07 & 0.16 & 0.17 & 0.50\tabularnewline
 & Insured Partly at fault $75\%$ & -0.55 & -0.27 & 0.61 &  -8.55 & 0.88 & 0.34 & 0.93 & -6.36 & 0.61 & 0.67 & 2.64 & -3.71 & 0.80 & 1.49 \tabularnewline
  & No fault & 4.30 & 3.23 & -1.19 & -5.06 & -1.25 & 0.05 & 6.37 & 1.61 & 2.30 & 4.04 & 1.29 & -2.13 & -0.83 & 0.87 \tabularnewline
  & Not applicable & 1.46 & -0.07 & -1.88 & -28.36 & -3.83 & -4.53 & 1.05 & -1.28 & -1.12 & -1.36 & -0.77 & -2.98 & -2.35 & -3.16 \tabularnewline
  & Not Available & 5.81 & 1.34 & 3.74 & -12.24  & 4.27 & 1.26 & 7.88 & 0.74 & 4.41 & 0.99 & 4.16 & -0.43 & 3.32 & 1.06 \tabularnewline
\bottomrule
\end{tabularx}}
\end{center}
\end{sidewaystable}

\newpage
\section{Choice of the severity models}
\label{ap:aic_bic}
\begin{table}[h]
\begin{center}
\caption{Choice of distributions for the severity}
\label{tab:modelselection}
\begin{tabular}{@{}l l r r  @{}}
\toprule
\multirow{2}{*}{Coverage} & \multirow{2}{*}{Model} & \multicolumn{2}{c}{$\forall j$}  \\
  &  & \multicolumn{1}{c}{AIC} & \multicolumn{1}{c}{BIC} \\
\midrule
 \multirow{5}{*}{Accident Benefits} & Log-Normal & 1,101,322 & 1,101,790 \\
 & Gamma & 1,099,669 & 1,100,137 \\
 & Pareto & 1,093,130 & 1,093,598 \\
 & Generalized Beta II & \textbf{1,092,513} & \textbf{1,092,998} \\
 & Weibull & 1,096,566 & 1,097,034 \\
  \midrule
 \multirow{5}{*}{Bodily Injury} & Log-Normal & 519,469 & 519,896 \\
 & Gamma & 519,451 & 519,878 \\
 & Pareto & 520,152 & 520,580 \\
  & Generalized Beta II & 518,800 & 519,236 \\
 & Weibull & \textbf{517,012} & \textbf{517,439} \\
 \midrule
 \multirow{5}{*}{Vehicle Damage} & Log-Normal & 6,818,688 & 6,819,250 \\
 & Gamma & 6,811,968 & 6,812,530 \\
 & Pareto & 6,767,629 & 6,768,192 \\
  & Generalized Beta II & \textbf{6,760,099} & \textbf{6,760,684} \\
 & Weibull & 6,797,965 & 6,798,528 \\
 \midrule
 \multirow{5}{*}{Loss of Use} & Log-Normal & 2,318,483 & 2,319,003 \\
 & Gamma & 2,315,044 & 2,315,564\\
 & Pareto & 2,345,577 & 2,346,097  \\
  & Generalized Beta II & \textbf{2,305,159} & \textbf{2,305,699}\\
 & Weibull & 2,338,368 & 2,338,888\\
 \bottomrule
\end{tabular}
\end{center}
\end{table}

\newpage
\section{Results stability}
\label{ap:stability}
\begin{figure}[h!]
\centering
\includegraphics[width=1\textwidth]{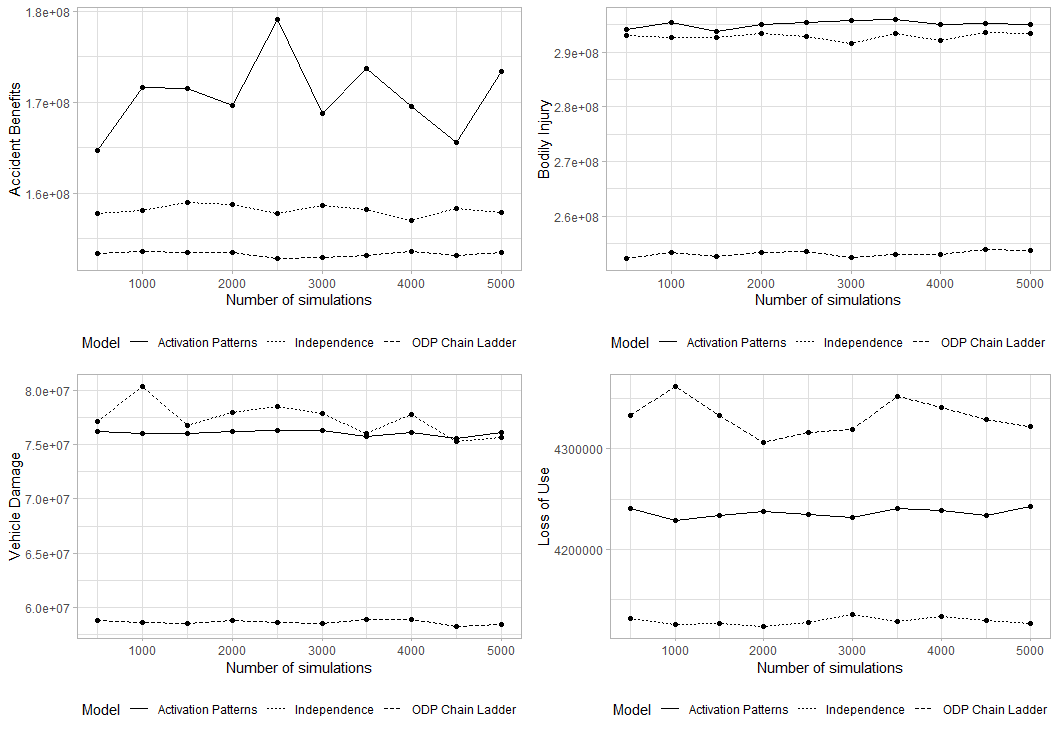}
\caption{Results of the RBNS reserves ($95\%$ VaR) based on the number of simulations performed}
\label{fig:Fig12}
\end{figure}

\clearpage

\bibliographystyle{unsrt}  
\bibliography{main}

\end{document}